\documentclass[twocolumn,nofootinbib,preprintnumbers,superscriptaddress,
amsmath,amssymb,prc]{revtex4-2}
\usepackage{amsmath}
\usepackage{color}
\usepackage{amsfonts}
\usepackage{graphicx}
\usepackage[utf8]{inputenc}
\usepackage{tabularx}
\usepackage{booktabs}
\usepackage{multirow} 
\usepackage{threeparttable} 
\usepackage{newtxtext,newtxmath}
\usepackage{caption}
\usepackage{subcaption}
\usepackage{float}
\usepackage{setspace}
\usepackage{geometry}
\usepackage{booktabs}  
\usepackage{multirow}  

\begin{document}

\title{$g_{\rm A}$-sensitive $\beta$ spectral shapes in the mass $A=86-99$ region assessed by the nuclear shell model}

\author{Marlom Ramalho}\thanks{Corresponding authors: madeoliv@jyu.fi}
\affiliation{Department of Physics, University of Jyv\"askyl\"a, P.O. Box 35, FI-40014, Jyv\"askyl\"a, Finland}

\author{Jouni Suhonen}\thanks{jouni.t.suhonen@jyu.fi}
\affiliation{Department of Physics, University of Jyv\"askyl\"a, P.O. Box 35, FI-40014, Jyv\"askyl\"a, Finland}
\affiliation{International Centre for Advanced Training and Research in Physics (CIFRA), P.O. Box MG12, 077125 Bucharest-Magurele, Romania}

\begin{abstract}
Recent years have witnessed an expanding interest in experimental studies of $\beta$ electrons (electrons emitted in $\beta^-$ decay transitions) and their energy distributions, the so-called $\beta$-electron spectra. These experiments are interested mainly in $\beta$ transitions with electron spectra sensitive to the effective value of the weak axial coupling $g_{\rm A}$. In the present paper we make an extensive search for $g_{\rm A}$ sensitive $\beta$ spectral shapes in the $A=86-99$ region using the nuclear shell model with the well established Hamiltonians \textit{glekpn} and \textit{jj45pnb}, designed to render a good description of the spectroscopic properties of nuclei in this mass region. We have found eight $\beta^-$ decay transitions with various degrees of $g_{\rm A}$ sensitivity. Moreover, these transitions are also important in pinning down the value of the so-called small relativistic vector nuclear matrix element, sNME. In addition, some of the corresponding mother nuclei are important contributors to the antineutrino flux from nuclear reactors. All this means that the found 
$\beta$ transitions are potentially of great interest for future rare-events experiments.
\end{abstract}
\maketitle

\section{Introduction}

Rare-events experiments typically look for beyond-the-standard-model (BSM) physics by, e.g., measurements of rare nuclear $\beta$ decays and double $\beta$ decays. Lately, a booming interest in these experiments has concentrated on studies of $\beta$ electrons (electrons emitted in $\beta^-$ decays) and their energy distributions, the so-called 
$\beta$-electron spectra. Experimental and/or theoretical information on these spectra is crucial for, e.g., resolving the anomalies related to the antineutrino flux from nuclear reactors \cite{Hay2019,Ram2022}, and for resolving the common backgrounds in the rare-events experiments themselves \cite{Ram2023}. Also, pinning down the effective values of weak couplings is a considerable incentive for those present and future $\beta$-decay experiments able to tackle the spectral shapes of $\beta$ electrons. In particular, determination of the effective value of the weak axial coupling $g_{\rm A}$ is needed for the estimation of the sensitivities of experiments trying to measure the neutrinoless double beta ($0\nu\beta\beta$) decay since the $0\nu\beta\beta$ half-life is proportional to the inverse 4$th$ power of 
$g_{\rm A}$ \cite{Suh2017,Suh17-rev,Suh19-rev}. The implications of detection of this decay mode are fundamental as discussed in the recent reviews \cite{Eng2017,Eji2019,Bla2020,Ago2023}. 

Nuclear $\beta$ decays vary in complexity, from allowed to highly forbidden ones: In the allowed Fermi and Gamow-Teller decays no orbital angular momentum is transferred to the emitted leptons \cite{Suh2007}, whereas in the forbidden decays \cite{Beh1982} a non-zero orbital angular momentum is transferred. Like in the case of allowed $\beta$ decays, also in the case of forbidden unique $\beta$ decays the lepton phase space can be separated from the nuclear part, resulting in a universal $\beta$ spectral shape, independent of nuclear-structure details \cite{Suh2007}. Of special interest for the rare-events experiments are the 
forbidden non-unique $\beta$ decays for which the $\beta$ spectral shapes can be strongly nuclear-structure dependent through several nuclear matrix elements (NME). These NME depend, in turn, on the structure of the wave functions of the states involved in a $\beta$-decay transition.

In addition to the many NME, the (partial) half-life of a forbidden non-unique $\beta$ transition depends on the so-called effective value of $g_{\rm A}$, recently discussed in the reviews \cite{Eji2019,Suh17-rev,Suh19-rev}. As discussed in these reviews, the effective value of $g_{\rm A}$ is quenched relative to the bare-nucleon value $g_{\rm A}=1.27$. Only in rare cases, for first-forbidden $\beta$-decay transitions with change in parity and no change in angular momentum, there is an enhancement of $g_{\rm A}$ present \cite{Kos2018}.
As mentioned earlier, pinning down the effective value of $g_{\rm A}$ is of crucial importance for quantifying the impact on the BSM physics stemming from the gained data of rare-events experiments. As discussed in \cite{Haa2016,Haa2017}, this information can be gained by using the so-called spectrum-shape method (SSM). Use of SSM requires a $\beta$-electron spectrum with a notable $g_{\rm A}$ dependence in its shape. In this case information on the effective value of $g_{\rm A}$ can be gained through the comparison of computed template $\beta$ spectra, for different $g_{\rm A}$ values, with the measured one in the aim to find a match.
Such SSM analyses of $\beta$-spectral shapes of individual $\beta^-$ transitions have been done recently for the fourth-forbidden non-unique $\beta$ decays of $^{113}$Cd and $^{115}$In in \cite{Bod2020,Kos2021,Led2022}. An enhanced version of SSM (enhanced SSM) was introduced in \cite{Kum2020,Kum2021} and the spectral moments method (SMM) in \cite{Kos2023}. Measurements of the $^{113}$Cd and $^{115}$In $\beta$ spectra are being extended also to other potentially sensitive candidates, like in the case of the ACCESS Collaboration \cite{Pag2023}.

An additional ingredient in the theoretical analyses of $\beta$ spectral shapes is the so-called small relativistic NME, the sNME, used to fix the measured (partial) half-life in the enhanced SSM \cite{Kum2020,Kum2021} and the SMM \cite{Kos2023}. In spite of its smallness, sNME can influence the (partial) half-lives and shapes of $\beta$-electron spectra quite strongly, see \cite{Kos2021,Kos2023,Ram2023}. The sNME gathers its major contributions outside the proton (neutron) valence major shell that contains the proton (neutron) Fermi surface. This makes its calculation particularly hard for the nuclear shell model (NSM), exploited in this work, which typically uses as valence space just the valence major shell for both protons and neutrons. 

The sNME can be related to the so-called large vector NME, l-NME, by using the CVC (conserved vector current) hypothesis \cite{Beh1982}. The l-NME gathers its major contributions from the valence major shells so that it is reliably calculable in the framework of the NSM. Although this CVC-dictated value of the sNME is an idealization, strictly applicable to an ideal nuclear many-body calculation \cite{Beh1982}, it still serves as a good reference in our search for a realistic value of the sNME. In our present work we determine the values of the sNME by fitting the experimental partial half-lives (branching ratios) corresponding to the eight $\beta$-decay transitions that we have found to depend on the values of $g_{\rm A}$ and/or sNME. In this case, the dependence of the $\beta$ spectral shape on the value of sNME stems from the fact that there are always two values (or none at all) of the sNME that reproduce the measured branching of a $\beta$ transition.

The present article is organized as follows: In Sec.II the adopted theoretical framework is summarized by introducing briefly the $\beta$ spectral shapes and the related NME, as also their computation using the NSM. Our results are presented and discussed in Sec. III. Conclusions are drawn in Sec. IV.

\section{Theoretical framework}

In this work we discuss only $\beta^-$ decays which are weak-interaction processes where a neutron transmutes into a proton and an electron and an electron antineutrino ($\bar{\nu}_e$) are emitted within a nuclear environment:
\begin{equation*}
n \rightarrow p + e^- + \bar{\nu}_e.
\end{equation*}

In the following, we describe briefly the theory of $\beta$-electron spectral shapes, and their connection to the effective value of the weak axial coupling $g_{\rm A}$. We also discuss the NME, in particular the sNME and its role in the present calculations. 
Furthermore, the NSM, alongside its effective interactions and single-particle model spaces are discussed.

\subsection{$\beta$ spectral shapes}

The partial half-life corresponding to a branching ratio of a transition to a particular final state in the daughter isobar can be obtained from the expression
\begin{equation}
t_{1/2} = \kappa/\tilde{C} \,,
\end{equation}
where $\kappa=6289\,\textrm{s}$ is a collection of natural constants \cite{Kum2021} and the integrated shape function reads
\begin{equation}
\tilde{C} = \int_0^{w_0}S(w_e)dw_e \,.
\label{eq:cee}
\end{equation}
where the shape function $S(w_e)$ can be written as
\begin{equation}
S(w_e) = C(w_e)pw_e(w_0-w_e)^2F_0(Z,w_e) \,.
\label{eq:cee}
\end{equation}
In this expression, $F_0(Z,w_e)$, with $Z$ as the proton number of the daughter nucleus, is the usual Fermi function taking into account the final-state Coulomb distortion of the wave function of the emitted electron and 
\begin{equation}
w_0 = \frac{W_0}{m_e c^2}, \quad w_e = \frac{W_e}{m_e c^2}, \quad p = \frac{p_e c}{m_e c^2} = \sqrt{w_e^2 - 1}
\end{equation}
are kinematic quantities scaled dimensionless by the electron rest mass $m_ec^2$. Here $p_e$ and $W_e$ are the momentum and energy of the emitted electron, respectively, and $W_0$ is the beta endpoint energy, which for the ground-state transitions defines the $\beta$-decay $Q$ value. The shape factor $C(w_e)\approx 1$ for allowed transitions \cite{Suh2007} and in general it is a complicated combination of leptonic phase-space factors and NME, as described in detail in \cite{Beh1982} and recently in \cite{Haa2016,Haa2017}.

In the current work, we discuss first-forbidden and second-forbidden non-unique $\beta^-$-decay transitions and the associated $\beta$-spectral shapes $S(w_e)$. The presently discussed first-forbidden $\beta$ transitions are pseudovector (change in parity) $\Delta J = 1$ (change of one unit of angular momentum) transitions, and the second-forbidden transition, corresponding to the decay of $^{99}$Tc, is a $\Delta J = 2$ tensor (no change in parity) transition. All these transitions have both vector and axial-vector components and depend on more than one NME, thus being sensitive to details of nuclear structure through the initial and final nuclear wave functions. For the vector part we adopt the CVC-compatible value $g_{\rm V}=1.0$ of the weak vector coupling. In this work, we will refer to the effective $g_{\rm A}^{\rm eff}$ as simply $g_{\rm A}$ and it should not be confused with its free-nucleon value $g_{\rm A}^{\rm free}=1.27$.

Our particular aim is to find $\beta$ spectral shapes that are sensitive to the effective value of 
$g_{\rm A}$. This dependence is enabled by the interference of the vector and axial-vector parts with the mixed vector-axial-vector part in the decomposition
\begin{equation}
S(w_e) = g_{\rm V}^2S_{\rm V}(w_e) +  g_{\rm A}^2S_{\rm A}(w_e) + g_{\rm V}g_{\rm A}S_{\rm VA}(w_e)
\label{eq:shape}
\end{equation}
of the shape function. Thus far only very few cases with a sizable sensitivity of $S(w_e)$ to the value of $g_{\rm A}$ have been identified \cite{Eji2019,Kum2020,Kum2021}. An other dependence of $S(w_e)$ can come from the chosen value of sNME \cite{Ram2023}. Details related to this chosen value of sNME are highlighted below.

\subsection{$\beta$ spectral shapes and the value of the sNME}

The small relativistic NME, sNME, can play an important role in combined studies of beta spectral shapes and branching ratios (partial half-lives) \cite{Kum2021,Kos2021,Kos2023,Ram2023}. In these works the sNME has been used as a fitting parameter, together with $g_{\rm A}$, in order to fit the experimental $\beta$ spectral shapes and branching ratios simultaneously. In the nuclear-structure calculations, the sNME gathers contributions outside the nucleon major shell(s) where the proton and neutron Fermi surfaces lie. Due to the limitation of the NSM valence space to these shells only, the value of the sNME turns out to be unrealistic (practically zero) in the NSM calculations. 

In an ideal case (infinite valence spaces, perfect nuclear many-body theory) the value of the sNME is tied to the value of the so-called large vector NME, l-NME, by the CVC (Conserved Vector Current) hypothesis \cite{Beh1982} through the relation
\begin{equation}
\label{eq:CVC-sNME}
   ^V\mathcal{M}_{KK-11}^{(0)} = \left(\frac{\frac{{(-M_n c^2 + M_p c^2 + W_0) \times R}}{{\hbar c}} + \frac{6}{5} \alpha Z}{\sqrt{K (2K + 1)} \times R}\right) \times {}^V\mathcal{M}_{KK0}^{(0)} \,,
\end{equation}
where the left side of the equation is the sNME, the last term on the right is the l-NME, and $K$ denotes the order of forbiddenness, with $K=1$ ($K=2$) denoting the first-forbidden (second-forbidden) decays. The quantities $M_n$ and $M_p$ denote neutron and proton masses, respectively. $W_0$ is the available endpoint energy for the decay,  $\hbar$ the reduced Planck constant, $\alpha$ is the fine-structure constant, and $c$ the speed of light. Lastly, $Z$ is the atomic number of the daughter nucleus, and  $R = 1.2 A^{1/3}$ is the nuclear radius in fm \cite{Suh2007}, $A$ being the nuclear mass number. 
The value of the l-NME can be rather reliably computed by the NSM since the main contributions to it stem from the major shell(s) where the nucleon Fermi surfaces lie. The CVC value of sNME can thus be considered as a good reference for the proper value of the sNME.

In our calculations, we adopt the approach of fitting the sNME such that each individual $\beta^-$ transition can be reproduced in terms of the branching ratio (partial half-life) accounting for screening, radiative, and atomic exchange corrections. Visible at low electron energies is the atomic exchange correction which was originally derived for allowed $\beta$ decays \cite{Nitescu2023} and is responsible for the upward tilt seen in all curves. The experimental branching ratios are taken from \cite{ENSDF}.
There is a quadratic dependence of the computed branching ratios on the value of the sNME and hence two values of the sNME, for each decay transition, reproduce the experimental branching (in some cases there are only complex-conjugate pairs of solution available, meaning that the experimental branching cannot be reproduced by the adopted NSM Hamiltonian). One of these two sNMEs is closer to the CVC value of the sNME and thus offers a way to define the ``optimal'' beta spectral shape: By this hypothesis, choosing always the sNME closer to its CVC value produces the most probable spectral shape for a given $\beta$-decay transition. In the following we study how clear is this selection of the "closer-to-the-CVC" value of the sNME, and the dependence of the $\beta$ spectral shape on this chosen value.



\subsection{Nuclear Shell-Model calculations}

\begin{table}[ht]
\centering
\caption{Single-particle valence spaces and single-particle energies adopted in the present work. The mass ranges are: Set 1 is for $A < 88$, Set 2 is for $A=88-98$, and Set 3 is for $A=94-98$.}
\label{table:transposed}
\begin{tabular}{cccc|c}
\toprule
& \multicolumn{3}{c|}{glekpn (MeV)} & \multirow{2}{*}{ jj45pnb (MeV)} \\
\cmidrule(lr){2-4}
& Set 1 & Set 2 & Set 3 & \\
\midrule
$\pi$0f$_{7/2}$ & -10.480 & -8.980 & -8.012 & - \\
$\pi$0f$_{5/2}$ & -5.678 & -4.178 & -4.197 & -14.938 \\
$\pi$1p$_{3/2}$ & -5.761 & -4.261 & -2.796 & -13.437 \\
$\pi$1p$_{1/2}$ & -1.693 & -1.693 & -1.340 & -12.0436 \\
$\pi$0g$_{9/2}$ & -1.423 & -1.423 & -0.436 & -8.9047 \\
\midrule
$\nu$0g$_{9/2}$ & -9.306 & -9.306 & -10.357 & - \\
$\nu$0g$_{7/2}$ & 10.927 & 10.927 & 11.622 & 6.2302 \\
$\nu$1d$_{5/2}$ & 4.220 & 4.220 & 5.236 & 2.4422 \\
$\nu$1d$_{3/2}$ & 7.212 & 7.217 & 9.496 & 2.9448 \\
$\nu$2s$_{1/2}$ & 4.371 & 4.371 & 6.710 & 2.6738 \\
$\nu$0h$_{11/2}$ & - & - & - & 4.3795 \\
\bottomrule
\end{tabular}
\end{table}

The NSM calculations were performed using the software $\textit{KSHELL}$~\cite{Shi2019} with the Hamiltonians \textit{glekpn}~\cite{Mach1990} and \textit{jj45pnb}~\cite{Lisetskiy2004}.
We have used two different single-particle valence spaces for \textit{glekpn} and \textit{jj45pnb}, as indicated in Table~\ref{table:transposed}. In particular, for the \textit{glekpn} valence space we have employed three different single-particle energy sets, such that Set 1 is suited for the masses 
$A < 88$,  Set 2 for the mass range $A=88-98$, and Set 3 for the mass range $A=94-98$. Set 1 is an adjustment of Set 2 changing only the first three proton single-particle energies.

For all \textit{glekpn} computations the valence spaces have been truncated by including the $\pi$0f$_{7/2}$ orbital to be part of the closed core, and the same is true for the $\nu$0g$_{9/2}$ orbital. In the case of the \textit{jj45pnb} calculations, due to $M$-scheme dimensions being in some cases higher than $10^{10}$, a truncation was made for $A=95$ such that the $\pi$0f$_{5/2}$ orbital was forced to have from 4 to 6 protons, and lastly, for $A=97$ the truncation consists of allowing up to 6 neutrons in the $\nu$0h$_{11/2}$ orbital and the orbital $\pi$0f$_{5/2}$ was forced to have from 3 to 6 protons. For all other masses, no truncation was made in the \textit{jj45pnb} calculations.

\section{Results}

Here we detail the steps involved in our calculations. First we discuss the electromagnetic observables of the involved states in the light of experimental data. Then the experimental branching ratios of the $\beta$-decay transitions of interest are fitted by varying the value of the small vector NME, sNME, for each selected value of the axial-vector coupling $g_{\rm A}$. We then plot the corresponding $\beta$ spectral shapes to see if there is any sensitivity of the shapes to the value of sNME and/or $g_{\rm A}$.

\begin{figure}[h]
\includegraphics[width=1.0\columnwidth]{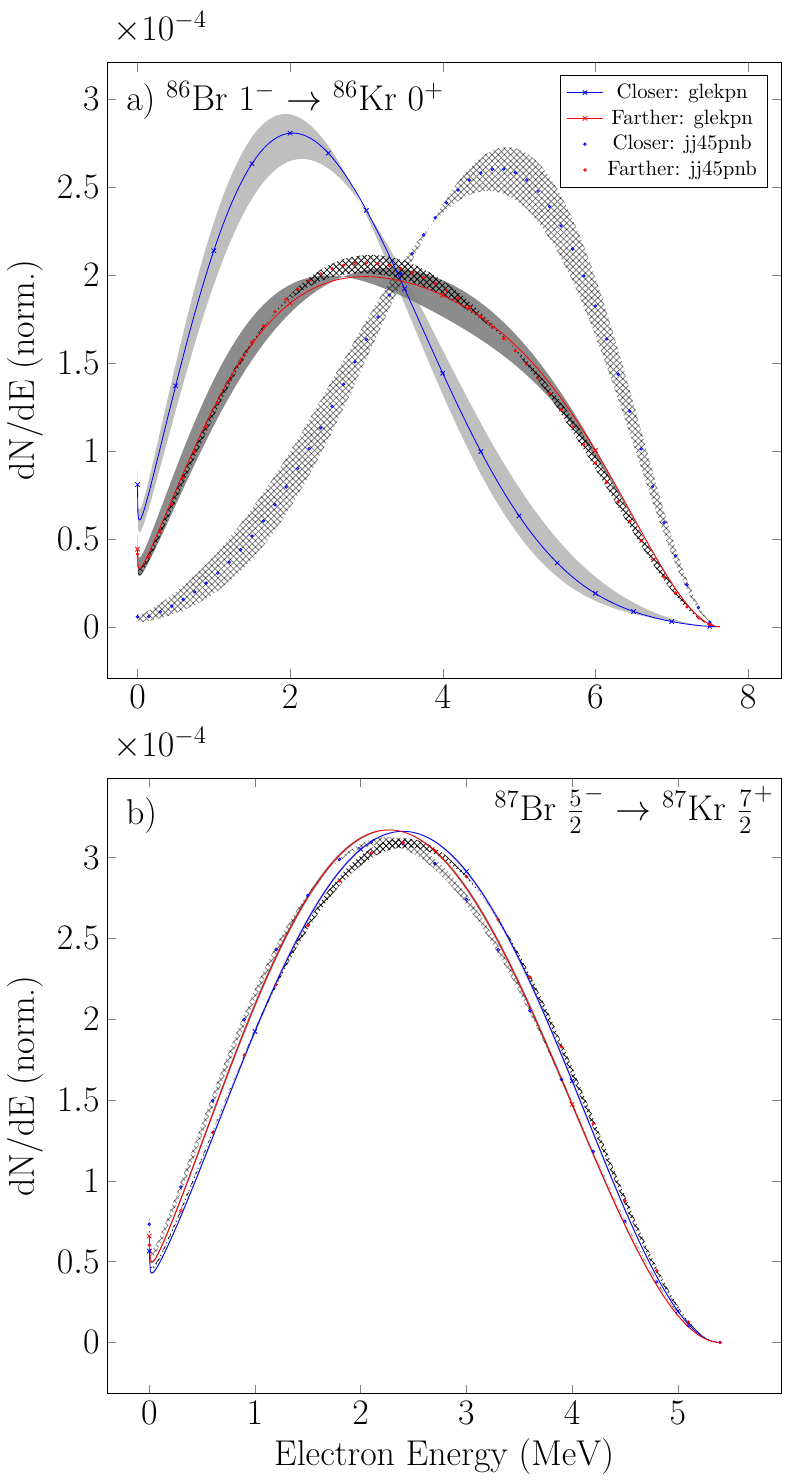}
\caption{\label{fig:86_87Br} Computed $\beta$ spectral shapes of the transitions $^{86}\textrm{Br}(1^-)\to\,^{86}\textrm{Kr}(0^+)$ [panel a)] and $^{87}\textrm{Br}(5/2^-)\to\,^{87}\textrm{Kr}(7/2^+)$ [panel b)]. The crossed-blue(red) curves are those constructed by adopting the value closer to (farther from) to the CVC value of the sNME for each transition. The light (darkened) gray-hatched regions denote the span of the curves corresponding to the range of $g_{\rm A}=0.8-1.2$ and their corresponding closer (farther) sNMEs 
for the \textit{glekpn} Hamiltonian. 
The corresponding blue and red dotted curves and their light (darkened) gray-hatched regions show the results for the \textit{jj45pnb} interaction. All the areas under the curves have been normalized to unity.}
\end{figure}

\subsection{Electromagnetic observables}

\begin{table*}[ht]
\centering
\caption{Comparison of the experimental and \textit{jj45pnb}-computed and \textit{glekpn}-computed state energies $E_{\rm exc}$ (in units of MeV), electric quadrupole moments $Q$ (in units of barn), and magnetic dipole moments $\mu$ (in units of nuclear magneton $\mu_N$). The experimental values are taken from the evaluation \cite{ENSDF}. The values in squared brackets denote which set of single-particle energies of \textit{glekpn} was used in the calculation. The adopted effective charges are $e_{\rm eff}^p$ = 1.5e and $e_{\rm eff}^n$ = 0.5e and the bare g-factors $g_l(p)$ = 1, $g_l(n)$ = 0, $g_s(p)$ = $5.585$, and $g_s(n)$ = $-3.826$ were used for the magnetic moments.}
\label{table:EM-observables}
\begin{tabular}{lclr|ccc|ccc}
\toprule
\multicolumn{4}{c|}{Experimental Evaluation} & \multicolumn{3}{c|}{jj45pnb} & \multicolumn{3}{c}{glekpn} \\
\cmidrule(lr){1-4} \cmidrule(lr){5-7} \cmidrule(lr){8-10}
Isotope ($J^\pi$) & $E_{\rm exc}$ (MeV) & $Q$ (barn) & $\mu$ ($\mu_N$) & $E_{\rm exc} $ (MeV) & $Q$ (barn) & $\mu$ ($\mu_N$) & $E_{\rm exc}$ (MeV)& $Q$ (barn) & $\mu$ ($\mu_N$) \\
\midrule
$^{86}$Br $(1^{-})$ & 0.000 & - & - & 0.601 & +0.079 & +1.929 & 0.614 & -0.034 & -0.455 [1] \\
$^{86}$Kr $(2^{+})$ & 1.565 & - & +2.20(10) & 1.614 & -0.148 & +2.084 & 1.585 & -0.123 & +2.984 [1] \\
$^{86}$Kr $(4^{+})$ & 2.250 & - & +4.1(6) & 2.265 & +0.450 & +3.966 & 2.340 & +0.347 & +3.724 [1] \\
$^{87}$Br $(\frac{5}{2}^{-})$ & 0.000 & - & - & 0.519 & -0.016 & +2.024 & 0.000 & +0.376 & +0.612 [1] \\
$^{87}$Kr $(\frac{7}{2}^{+})$ & 1.420 & - & - & 1.532 & -0.024 & -0.003 & 1.707 & -0.106 & +1.124 [1] \\
$^{87}$Kr $(\frac{5}{2}^{+})$ & 0.000 & -0.300(3) & -1.022(2) & 0.000 & -0.295 & -1.683 & 0.000 & -0.188 & -1.683 [1] \\
$^{87}$Kr $(\frac{5}{2}^{+})$ & 0.000 & -0.300(3) & -1.022(2) & 0.000 & -0.295 & -1.683 & 0.000 & -0.359 & -1.408 [2] \\
$^{87}$Rb $(\frac{3}{2}^{-})$ & 0.000 & +0.1335(5) & +2.75129(8) & 0.000 & +0.166 & +2.842 & 0.658 & +0.177 & +3.255 [2] \\
$^{93}$Y $(\frac{1}{2}^{-})$ & 0.000 & - & -0.139(1) & 0.000 & - & -0.266 & 0.747 & - & -0.538 [2] \\
$^{93}$Zr $(\frac{3}{2}^{+})$ & 0.267 & - & - & 0.198 & +0.001 & -0.042 & 0.183 & -0.108 & +0.637 [2] \\
$^{95}$Sr $(\frac{1}{2}^{+})$ & 0.000 & - & -0.537(2) & 0.182 & - & -1.650 & 0.275 & - & -0.751 [2] \\
$^{95}$Y $(\frac{1}{2}^{-})$ & 0.000 & - & -0.16(3) & 0.000 & - & -0.286 & 0.646 & - & -0.535 [2] \\
$^{95}$Y $(\frac{3}{2}^{-})$ & 0.686 & - & - & 0.308 & +0.279 & +1.700 & 0.000 & +0.396 & +2.146 [2] \\
$^{97}$Zr $(\frac{1}{2}^{+})$ & 0.000 & - & -0.936(5) & 0.000 & - & -1.829 & 0.271 & - & -0.757 [2] \\
$^{97}$Nb $(\frac{3}{2}^{-})$ & 1.251 & - & - & 0.925 & +0.272 & +1.809 & 0.000 & +0.442 & +2.174 [2] \\
$^{99}$Mo $(\frac{1}{2}^{+})$ & 0.000 & - & +0.375(3) & 0.202 & - & -1.799 & 0.639 & - & -0.835 [3] \\
$^{99}$Mo $(\frac{5}{2}^{+})$ & 0.098 & - & -0.775(5) & 0.222 & +0.435 & -0.266 & 0.000 & +0.500 & +1.062 [3] \\
$^{99}$Tc $(\frac{3}{2}^{-})$ & 0.509 & - & - & 0.696 & +0.020 & +1.825 & 0.763 & +0.350 & +2.255 [3] \\
$^{99}$Tc $(\frac{9}{2}^{+})$ & 0.000 & -0.129(6) & +5.687(2) & 0.112 & +0.022 & +5.662 & 0.512 & +0.243 & +5.416 [3] \\
$^{99}$Tc $(\frac{7}{2}^{+})$ & 0.141 & - & +4.48(15) & 0.000 & -0.091 & +4.593 & 0.397 & +0.724 & +4.715 [3] \\
$^{99}$Tc $(\frac{5}{2}^{+})$ & 0.181 & - & +3.48(4) & 0.063 & -0.476 & +3.356 & 0.638 & +0.307 & +3.393 [3] \\
$^{99}$Ru $(\frac{5}{2}^{+})$ & 0.000 & +0.079(4) & -0.641(5) & 0.037 & -0.002 & +0.288 & 0.000 & -0.322 & +1.083 [3] \\
$^{99}$Ru $(\frac{3}{2}^{+})$ & 0.090 & +0.231(13) & -0.248(6) & 0.000 & +0.055 & -0.282 & 0.276 & +0.275 & +0.569 [3] \\
\bottomrule
\end{tabular}
\end{table*}

We probe the reliability of our adopted nuclear wave functions by first computing their electromagnetic properties, namely their electric quadrupole moments $Q$ in units of barn and their magnetic dipole moments $\mu$ in units of nuclear magneton $\mu_N$. We compare these computed values, as well as the computed energies, with the available data in Table~\ref{table:EM-observables}. Here the states are given in column 1 and their experimental excitation energies, quadrupole, and dipole moments in columns 2-4. The corresponding computed excitation energies, quadrupole, and dipole moments are given in colums 5-7 for the \textit{jj45pnb} Hamiltonian and in colums 8-10 for the \textit{glekpn} Hamiltonian.

From Table~\ref{table:EM-observables} one can see that mostly the computed energies of the states are in fair agreement with experiment. Since almost all these states are in odd-mass nuclei, the state density can be quite high even at low excitation energies. This makes the prediction of the correct ground state sometimes quite tricky for the NSM. This can be seen in Table~\ref{table:EM-observables} as a failure of the NSM to predict the correct ground state. On the other hand, this is not a serious flaw since in all these cases the experimentally determined ground state is not far in excitation in the predicted theoretical energy spectrum. 

In addition to the state energies, a measure of the quality of the corresponding computed wave functions, relevant for the present purposes, are the electromagnetic moments. From Table~\ref{table:EM-observables} one can see that the two interactions mostly agree in sign for both the electric quadrupole moments $Q$ and magnetic dipole moments $\mu$ of the states involved. The signs between the computed and measured moments differ only in cases where the absolute values of these moments are relatively small, like in the cases of the states $^{99}\textrm{Mo}(1/2^+,5/2^+)$ ($\mu$), $^{99}\textrm{Tc}(9/2^+)$ ($Q$), and $^{99}\textrm{Ru}(5/2^+)$ (both $\mu$ and $Q$). However, overall, the correspondence between the computed and measured values of these moments is quite good.

\subsection{Values of the small vector matrix element \label{subsec:sNME}}

\begin{table*}[ht]
\centering
\begin{threeparttable}
\caption{Ranges of the values of the fitted sNME (lower and upper bounds), corresponding to range $g_{\rm A}=0.8-1.2$ of the axial coupling, and the CVC value of the sNME for the Hamiltonians \textit{jj45pnb} (columns 6-8) and \textit{glekpn} (columns 9-11). The transition is given in columns 1 and 2, and the corresponding measured $Q$-value, excitation energy of the final state, and the branching are given in columns 3-5. The data are taken from the evaluation \cite{ENSDF}. The numbers without parentheses (in parentheses) correspond to the sNME range which is considered to be more (less) correlated with the CVC value of the sNME.}
\label{table:sNMEs}
\fontsize{8.8}{10}\selectfont
\begin{tabular}{ccccc|ccc|ccc}
\toprule
\multicolumn{5}{c|}{Evaluation Data} & \multicolumn{3}{c|}{jj45pnb ($\times 10^{-2}$)} & \multicolumn{3}{c}{glekpn ($\times 10^{-2}$)} \\
\cmidrule(lr){1-5} \cmidrule(lr){6-8} \cmidrule(lr){9-11}
$J_i^{\pi}$ & $J_f^{\pi}$ & $Q_{\rm exp}$ (MeV) & $E_{\rm exp}$ (MeV) & Br. (\%) & Lower & Upper & CVC & Lower & Upper & CVC \\
\midrule
$^{86}$Br $(1^{-})$ & $^{86}$Kr $(0^{+})$ & 7.633(3) & 0.000 & 15(8) & +1.55(-0.16) & +2.02(+0.13) & +4.99 & +3.40(+1.79) & +3.53(+2.33) & +5.36 \\
$^{87}$Br $(\frac{5}{2}^{-})$ & $^{87}$Kr $(\frac{7}{2}^{+})$ & 6.818(3) & 1.420 & 4.8(16) & +1.76(-1.18) & +1.83(-1.00) & +0.50 & -1.48(+1.71) & -1.43(+1.76) & +0.06 \\
$^{87}$Kr $(\frac{5}{2}^{+})$ & $^{87}$Rb $(\frac{3}{2}^{-})$ & 3.88827(25) & 0.000 & 30.5(22) & +0.57(-0.59) & +0.96(-0.57) & +1.57 & -1.20(+0.23) & -0.72(+0.29) & -2.79 \\
$^{93}$Y $(\frac{1}{2}^{-})$ & $^{93}$Zr $(\frac{3}{2}^{+})$ & 2.895(10) & 0.267 & 4.9(9) & -0.66(-0.68) & -0.08(-0.48) & +1.35 & -0.53(-0.20) & -0.50(-0.07) & -0.61 \\
$^{95}$Sr $(\frac{1}{2}^{+})$ & $^{95}$Y $(\frac{3}{2}^{-})$ & 6.090(7) & 0.686 & 8.9(7) & -2.04(+4.21) & -1.62(+4.55) & +1.04 & +3.64(+0.40) & +5.02(+3.64) & +6.82 \\
$^{97}$Zr $(\frac{1}{2}^{+})$ & $^{97}$Nb $(\frac{3}{2}^{-})$ & 2.659(2) & 1.251 & 3.90(20) & -0.10(-1.41) & +0.10(-1.22) & -0.52 & -1.77(-1.20) & -1.31(-0.80) & -3.70$^{a}$ \\
$^{99}$Mo $(\frac{1}{2}^{+})$ & $^{99}$Tc $(\frac{3}{2}^{-})$ & 1.3578(9) & 0.509 & 1.16(2) & -0.50(-1.38) & -0.23(-1.12) & -0.47 & -1.70(-0.96) & -1.66(-1.10) & -3.43 \\
$^{99}$Tc $(\frac{9}{2}^{+})$ & $^{99}$Ru $(\frac{5}{2}^{+})$ & 0.2975(10) & 0.000 & 99.9984(4) & -8.56(+4.96) & -5.15(+8.33) & -17.1 & -3.11(+10.4) & -0.93(+12.4) & +0.80 \\
\bottomrule
\end{tabular}
\begin{tablenotes}
\item[a] For this transition, the \textit{glekpn} Hamiltonian could not reproduce the measured branching ratio for $g_{\rm A}$ = 1.2 thus the ranges are for $g_{\rm A}$=$0.8-1.1$.
\end{tablenotes}
\end{threeparttable}
\end{table*}

As already mentioned above, we use the sNME as a fitting parameter to match, for each selected value of the axial coupling $g_{\rm A}$, the computed and measured branching ratios of the $\beta$-decay transitions of interest. We obtain two solutions for the value of sNME for each value of $g_{\rm A}$, giving two ranges of sNME values corresponding to our adopted range $g_{\rm A}=0.8-1.2$. These ranges are compared with the CVC value of the sNME, Eq.~(\ref{eq:CVC-sNME}), in Table~\ref{table:sNMEs}. In this table we give, for each individual $\beta$ transition (columns 1 and 2), the corresponding experimental $Q$-value and excitation energy in the final nucleus in units of MeV in columns 3 and 4. We also give the measured branching ratio in percents in column 5. The CVC values of the sNME are given in columns 8 and 11 for the Hamiltonians \textit{jj45pnb} and \textit{glekpn}, respectively. In columns 6 and 7 (columns 9 and 10) we give the lower and upper bounds of the range of the fitted sNME values, corresponding to the range $g_{\rm A}=0.8-1.2$ of the axial coupling, for the \textit{jj45pnb} (\textit{glekpn}) Hamiltonian. Here the numbers without parentheses (in parentheses) correspond to the range which is considered to be the more (less) compatible one with the CVC value of the sNME.

Taking a look at Table~\ref{table:sNMEs} indicates that there is a clear correlation of the ranges of the fitted sNME values with the CVC value of sNME for the decay of $^{87}$Kr for both Hamiltonians, for the decays of $^{97}$Zr and $^{99}$Mo for the \textit{glekpn} Hamiltonian, and the decays of $^{86}$Br, $^{95}$Sr, and $^{99}$Tc for the \textit{jj45pnb} Hamiltonian. In these cases the sign of the CVC value of the sNME clearly defines the preferred range of the fitted sNME values. The rest of the cases are less clear and the assignment of the closer-to-CVC-value range is almost a matter of taste, the decay of $^{93}$Y being the most unclear case. In the end, only experimental data on the $\beta$ spectral shapes, when compared with the corresponding computed shapes, will decide which range of the sNME values will be the more realistic one. 

\subsection{Beta spectral shapes}

\begin{figure*}[tbh] 
  \centering
  \begin{minipage}{0.49\textwidth}
    \includegraphics[width=\linewidth]{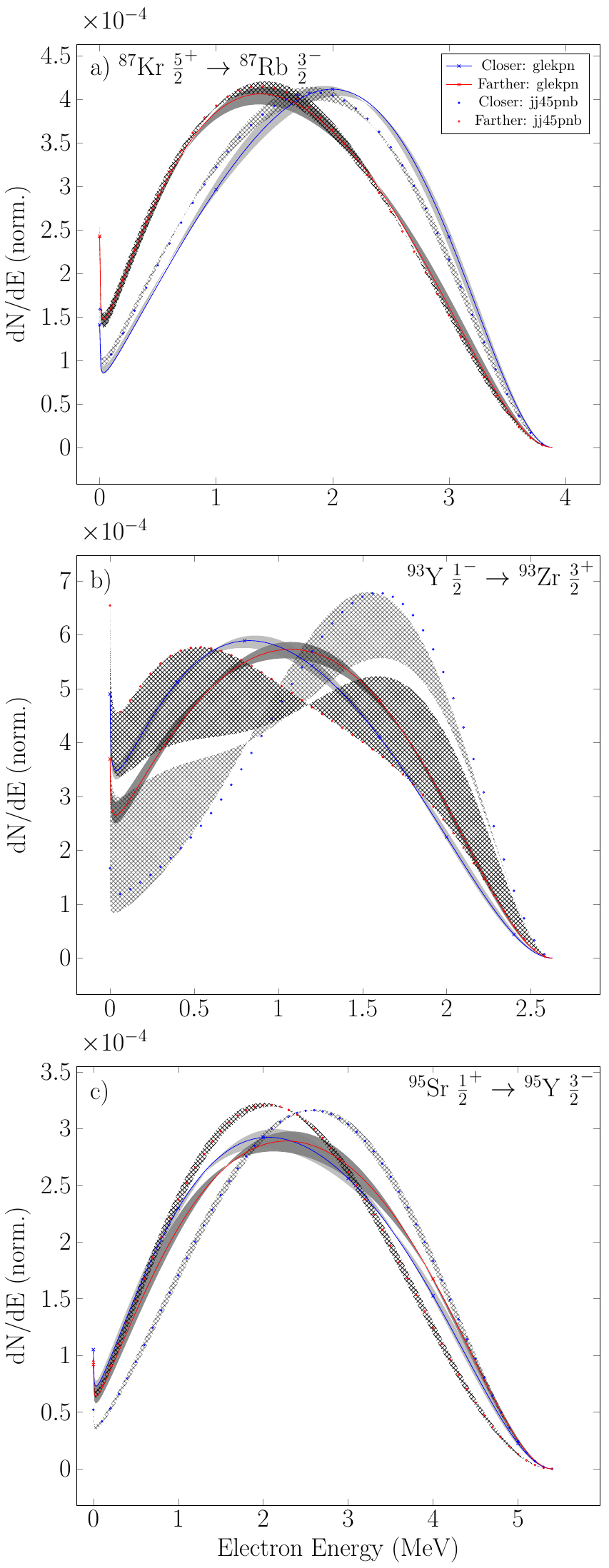}
    \caption{The same as in Fig.~\ref{fig:86_87Br} for the transitions $^{87}\textrm{Kr}(5/2^+)\to\,^{87}\textrm{Rb}(3/2^-)$ [panel a)], $^{93}\textrm{Y}(1/2^-)\to\,^{93}\textrm{Zr}(3/2^+)$ [panel b)], and $^{95}\textrm{Sr}(1/2^+)\to\,^{95}\textrm{Y}(3/2^-)$ [panel c)].}
    \label{fig:87_95Kr}
  \end{minipage}\hfill
  \begin{minipage}{0.49\textwidth}
    \includegraphics[width=\linewidth]{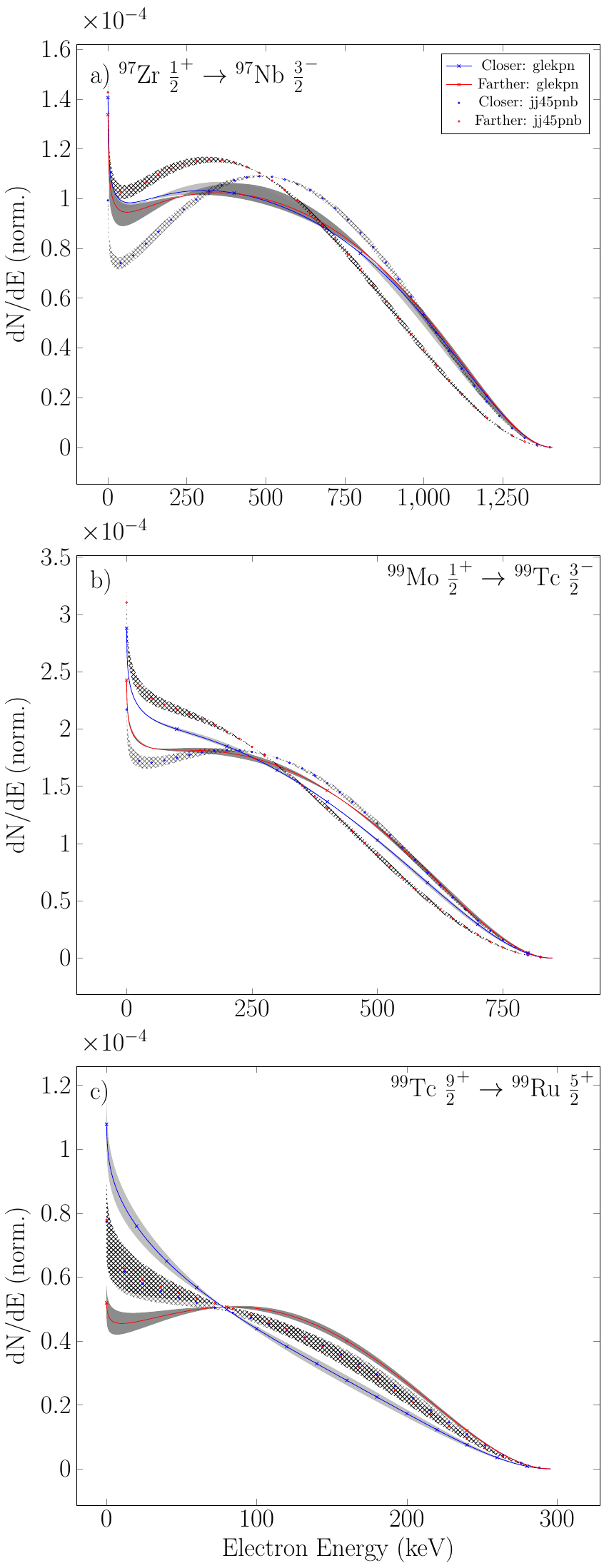}
    \caption{The same as in Fig.~\ref{fig:86_87Br} for the transitions $^{97}\textrm{Zr}(1/2^+)\to\,^{97}\textrm{Nb}(3/2^-)$ [panel a)], $^{99}\textrm{Mo}(1/2^+)\to\,^{99}\textrm{Tc} (3/2^-)$ ,[panel b)], and $^{99}\textrm{Tc}(9/2^+)\to\,^{99}\textrm{Ru}(5/2^+)$ [panel c)].}
    \label{fig:97_99Zr}
  \end{minipage}
\end{figure*}

We have produced the $\beta$ spectral shapes, corresponding to the shape function $S(w_e)$ of Eq.~(\ref{eq:cee}), by adopting the experimental $Q$-values and excitation energies listed in Table~\ref{table:sNMEs}. In addition, the experimental branching ratios of the table (column 5) have been reproduced by the sNME fitting procedure discussed in Sec.~\ref{subsec:sNME}. Our $\beta$-spectral results are summarized in Figs.~\ref{fig:86_87Br}, \ref{fig:87_95Kr}, and \ref{fig:97_99Zr}. In Fig.~\ref{fig:86_87Br} the decay transitions $^{86}\textrm{Br}(1^-)\to\,^{86}\textrm{Kr}(0^+)$ and $^{87}\textrm{Br}(5/2^-)\to\,^{87}\textrm{Kr}(7/2^+)$ are shown. It can clearly be seen that the decay of $^{86}\textrm{Br}$ depends very strongly on the adopted Hamiltonian, the axial coupling, and the sNME, whereas the decay of $^{87}\textrm{Br}$ depends quite weakly on all these three degrees of freedom. For the former decay the strong dependence on the used nuclear Hamiltonian is conspicuous for the closer-to-CVC values of the sNME, whereas there is practically no dependence on the chosen Hamiltonian for the farther sNME values.

In Fig.~\ref{fig:87_95Kr} the $\beta$ spectral shapes corresponding to the decay transitions $^{87}\textrm{Kr}(5/2^+)\to\,^{87}\textrm{Rb}(3/2^-)$, $^{93}\textrm{Y}(1/2^-)\to\,^{93}\textrm{Zr}(3/2^+)$, and $^{95}\textrm{Sr}(1/2^+)\to\,^{95}\textrm{Y}(3/2^-)$ are depicted. Here the decays of $^{87}\textrm{Kr}$ and $^{93}\textrm{Y}$ depend on the value of $g_{\rm A}$ and sNME, the latter even strikingly strongly. For the decay of $^{95}\textrm{Sr}$ there is only a very weak dependence on the value of $g_{\rm A}$ but a rather strong dependence on the value of the sNME, in particular for the Hamiltonian \textit{jj45pnb}. The decays of $^{93}\textrm{Y}$ and $^{95}\textrm{Sr}$ depend strongly on the chosen NSM Hamiltonian, whereas for the decay of $^{87}\textrm{Kr}$ there is practically no dependence on the chosen Hamiltonian.

In Fig.~\ref{fig:97_99Zr} the $\beta$ spectral shapes corresponding to the decay transitions $^{97}\textrm{Zr}(1/2^+)\to\,^{97}\textrm{Nb}(3/2^-)$, 
$^{99}\textrm{Mo}(1/2^+)\to\,^{99}\textrm{Tc} (3/2^-)$, 
and $^{99}\textrm{Tc}(9/2^+)\to\,^{99}\textrm{Ru}(5/2^+)$ are displayed. The decay of $^{97}\textrm{Zr}$ shows rather strong dependence on $g_{\rm A}$ for both Hamiltonians, but strong sNME dependence only for the \textit{jj45pnb} Hamiltonian. The dependence on the chosen Hamiltonian is notable. For the decay of $^{99}\textrm{Mo}$ there are strong dependencies on the sNME and the chosen Hamiltonian, whereas there is a notable $g_{\rm A}$ dependence only for the \textit{jj45pnb} Hamiltonian. In the case of the $^{99}\textrm{Tc}$ decay there is a strong dependence on the chosen Hamiltonian and the value of $g_{\rm A}$. The \textit{glekpn} interaction shows strong dependence on the values of sNME, whereas the \textit{jj45pnb} interaction shows only very moderate sNME dependence.

\section{Summary and conclusions}

In the present article we perform a survey of possible forbidden non-unique $\beta$-decay transitions which would be sensitive to the (effective) value of the weak axial-vector coupling $g_{\rm A}$. This dependence would allow determination of the value of this coupling in comparisons between the computed and measured electron spectral shapes, in terms of an enhanced spectrum-shape method (SSM) adopted in the present work. This enhanced method exploits the additional dimension of fitting the measured branching ratio of a $\beta$ transition by using the small relativistic vector NME, sNME, as a fitting parameter. 

Here we study the nuclear mass region $A=86-99$ since there are lots of possible decay transitions that have measured branching ratios that are reasonably large in order to enable realistic execution of $\beta$ spectral-shape measurements. In addition, in this mass region there are available two well-established Hamiltonians, \textit{jj45pnb} and \textit{glekpn} of the nuclear shell model (NSM), which would allow comparison of the results of these two Hamiltonians and a rough estimation of the uncertainties involved in our NSM calculations.

We have found 8 $\beta$-decay transitions of potential interest for spectral-shape measurements, the corresponding nuclei ranging from $^{86}$Br to $^{99}$Tc. The corresponding decay transitions can be grouped in four categories: 

\textbf{Category I} includes those transitions which are sensitive to the values of both $g_{\rm A}$ and sNME. These are the transitions $^{86}\textrm{Br}(1^-)\to\,^{86}\textrm{Kr}(0^+)$, $^{87}\textrm{Kr}(5/2^+)\to\,^{87}\textrm{Rb}(3/2^-)$, and $^{93}\textrm{Y}(1/2^-)\to\,^{93}\textrm{Zr}(3/2^+)$ for both Hamiltonians, and $^{97}\textrm{Zr}(1/2^+)\to\,^{97}\textrm{Nb}(3/2^-)$ and $^{99}\textrm{Mo}(1/2^+)\to\,^{99}\textrm{Tc}(3/2^-)$ for the \textit{jj45pnb} Hamiltonian, and $^{99}\textrm{Tc}(9/2^+)\to\,^{99}\textrm{Ru}(5/2^+)$ for the \textit{glekpn} Hamiltonian. 

\textbf{Category II} contains all $\beta$ transitions that have a strong $g_{\rm A}$ dependence but a weak sNME dependence. The corresponding transitions are $^{97}\textrm{Zr}(1/2^+)\to\,^{97}\textrm{Nb}(3/2^-)$ and $^{99}\textrm{Mo}(1/2^+)\to\,^{99}\textrm{Tc}(3/2^-)$ for the \textit{glekpn} Hamiltonian, and $^{99}\textrm{Tc}(9/2^+)\to\,^{99}\textrm{Ru}(5/2^+)$ for the \textit{jj45pnb} Hamiltonian.

\textbf{Category III} contains the $\beta$ transitions that are rather weakly sensitive to $g_{\rm A}$ but strongly sensitive to sNME. This one transition is $^{95}\textrm{Sr}(1/2^+)\to\,^{95}\textrm{Y}(3/2^-)$. 

\textbf{Category IV} includes those $\beta$ transitions that are rather weakly sensitive to both $g_{\rm A}$ and sNME. This one transition is $^{87}\textrm{Br}(5/2^-)\to\,^{87}\textrm{Kr}(7/2^+)$. 

All the discussed transitions are first-forbidden non-unique, except for the decay of $^{99}$Tc which is second-forbidden non-unique. These $\beta$ decay transitions are also associated to fission products that contribute notably to the antineutrino flux from nuclear reactors. In particular, the first-forbidden non-unique transitions may play a decisive role in solving the anomalies related to the reactor antineutrino flux \cite{Hay2019}.

The transition from \textbf{Category IV} is interesting mainly as a test of the nuclear-structure calculation, either proving or disproving the shape of the computed $\beta$ spectral shape. This philosophy is along the lines of the spectral-shape study of the $^{137}\textrm{Xe}(7/2^-)\to\,^{137}\textrm{Cs}(7/2^+)$ transition performed by the EXO-200 collaboration in \cite{EXO2020}. In this SSM study the measured and computed $\beta$ spectral shapes showed immaculate agreement, thus verifying the correctness of the corresponding nuclear-structure calculations, since the computed spectral shape of this first-forbidden non-unique transition turned out to be quite independent of the value of the axial-vector coupling. 

The transition from \textbf{Category III} is important in determining the proper value of the sNME, i.e. whether the fitted value closer to the CVC value of sNME is the right physical choice, as could be anticipated based on the \textit{jj45pnb} values of sNME in Table~\ref{table:sNMEs}. This same strategy is valid also for the $\beta$-decay transitions of \textbf{Category I}. In addition, the transitions of \textbf{Category I} and \textbf{Category II} open up a way to assess the effective value of the axial coupling, the ones of \textbf{Category II} even more straightforwardly. 

Building on the shape decomposition in Eq.~(\ref{eq:shape}), we can map the four categories to specific components within this equation: vector ($C_{\rm V}$), axial ($C_{\rm A}$), and vector-axial ($C_{\rm VA}$). \textbf{Category I} arises under two scenarios: first, when the vector-axial component, with a relevant sNME dependency, is dominant, resulting in a shape influenced by both $g_{\rm A}$ and sNME; second, when all three components have comparable magnitudes, and the sensitivity to the sNME in either (or both) the vector or vector-axial components leads to a similar dependence. For \textbf{Category II}, the dominance of the axial component or a dominant vector-axial component with weak sNME sensitivity both yield shapes primarily influenced by $g_{\rm A}$, albeit with slight sNME sensitivity. \textbf{Category III} is defined by a dominant vector component sensitive to sNME variations, making the shape dependent on sNME with minimal $g_{\rm A}$ influence. Finally, \textbf{Category IV} describes cases where a dominant and insensitive vector component to sNME variations results in a shape unaffected by both $g_{\rm A}$ and sNME.

Properties of many of the discussed $\beta$ spectral shapes depend more or less also on the adopted Hamiltonian, namely those corresponding to the decays of $^{93}$Y, $^{95}$Sr, $^{97}$Zr, $^{99}$Mo, and $^{99}$Tc, and also the one corresponding to the closer-to-CVC sNME for $^{86}$Br. These decay transitions then open up a way to also test the accuracy of the two widely used NSM Hamiltonians in describing also the $\beta$-decay properties in the nuclear-mass region of interest here.

Concerning the experimental aspects, the measurements of the presently discussed $\beta$ spectral shapes do not need very high precision in the very beginning owing to the large differences in many of the spectra with respect to the sNME and $g_{\rm A}$. A challenge for the scintillation-based experimental methods are the short half-lives of the decaying nuclei, excluding $^{99}$Tc, ranging from 55 seconds to 65.9 hours. Typically these methods are used for long-lived nuclei, like $^{113}$Cd \cite{Bel2007} and $^{115}$In \cite{Pfe1979}. Other possible methods are those based on semiconductor detectors \cite{Bod2020} and cryogenic calorimeters \cite{Led2022,Pag2023}. Short half-lives may be a challenge also for these type of measurement methods. Further methods are metallic magnetic calorimeters, the decay transition of $^{99}\textrm{Tc}(9/2^+)\to\,^{99}\textrm{Ru}(5/2^+)$ having already been measured by Paulsen \textit{et al.} \cite{Pau2023} by this method. The short half-lives may be overcome by using rediochemical methods or beams of radioactive nuclei \cite{Ten1989,Rud1990,Alg2021}, like in the ISOLDE facility at CERN. Interesting possibilities offers also the newly developed method employing the Ion Guide Isotope Separator On-Line facility at the Accelerator Laboratory of Jyv\"askyl\"a \cite{Gua2023}.

As a final note it should be stated that our calculations indicate that there are a lot of interesting possibilities for future $\beta$-decay experiments in the mass region $A=86-99$. These experiments would be in a position to shed light on the effective value of the weak axial coupling, on the role of the small relativistic vector matrix element in $\beta$-decay calculations, and on the capability of two well-established shell-model Hamiltonians in predicting $\beta$-decay properties, in addition to spectroscopic properties, of nuclei in this mass region.

\begin{acknowledgments}
We acknowledge the support by CSC – IT Center for Science, Finland, for the generous computational resources.
\end{acknowledgments}

\end{document}